\documentclass[conference]{IEEEtran}
\IEEEoverridecommandlockouts

\usepackage{cite}
\usepackage{amsmath,amssymb,amsfonts}
\usepackage{algorithmic}
\usepackage{graphicx}
\usepackage{textcomp}
\usepackage{xcolor}
\usepackage{multirow} 
\usepackage{siunitx} 
\def\BibTeX{{\rm B\kern-.05em{\sc i\kern-.025em b}\kern-.08em
    T\kern-.1667em\lower.7ex\hbox{E}\kern-.125emX}}
\begin{document}
\title{Efficient Realization of Multi-channel Visible Light Communication System for Dynamic Cross-Water Surface Channels\\}

\author{Han Qi, Tianrui Lin, Tianjian Wei, Qingqing Hu, Siyan Sun, Nuo Huang, and Chen Gong\\
	CAS Key Laboratory of Wireless-Optical Communications, School of Information Science and Technology,\\
	University of Science and Technology of China, Hefei 230027, China \\
	Email: qhjelena@mail.ustc.edu.cn, trlin@mail.ustc.edu.cn, weitj@mail.ustc.edu.cn, ruixihu@mail.ustc.edu.cn,\\  ssy22nb700@mail.ustc.edu.cn, huangnuo@ustc.edu.cn, cgong821@ustc.edu.cn
}
	\vspace{-0.5cm}

\maketitle

\begin{abstract}
This paper explores the transmission schemes for multi-channel water-to-air optical wireless communication (W2A-OWC) and introduces a prototype of a real-time W2A-OWC system based on a field-programmable gate array (FPGA). Utilizing an LED array as the transmitter and an APD array as the receiver, the system establishes a multi-channel transmission module. Such configuration enables parallel operation of multiple channels, facilitating the simultaneous transmission of multiple data streams and enhancing overall throughput. The FPGA serves as a real-time signal processing unit, handling both signal transmission and reception. By integrating low-density parity-check (LDPC) codes from 5G New Radio, the system significantly boosts its detection capabilities for dynamic W2A-OWC scenarios. The system also optimizes FPGA resource usage through time-multiplexing operation of an LDPC decoder's IP core. To evaluate the system's practicality, experiments were conducted under background radiation in an indoor water tank, measuring the frame error rate under both calm and fluctuating water surfaces. The experimental results confirm the feasibility and effectiveness of the developed W2A-OWC system.
\end{abstract}


\section{Introduction}
With an upsurge in underwater operations\cite{8936390,9241884,2019Recent}, the pursuit of seamless data exchange across aquatic boundaries has become a critical objective. To ensure effective data transfer, it is imperative to establish robust communication connecting land-based entities, such as communication hubs and aerial drones, with their submerged counterparts, including various sensors and underwater autonomous vehicles (UAVs)\cite{8121387},\cite{Xu:19}. 

Within the realm of mid-short range contactless communication, especially for applications that involve crossing water surfaces, the main communication carriers under consideration encompass acoustic waves, radio frequency (RF) waves, and optical waves \cite{7593257}. Acoustic waves, as a sophisticated choice for underwater communication, can achieve long transmission distances, potentially spanning several kilometers. However, they are hindered by limited bandwidth, resulting in low data rates and high latency, which makes them unsuitable for high-speed communication needs \cite{Zhou2019StudyOP}. Moreover, the water-air interface reflects a considerable portion of the acoustic energy, causing significant signal attenuation \cite{9601292}. 

On the other hand, RF signals are capable of traversing significant distances and sustaining high transmission rates in the air, with the potential to reach tens of kilometers and deliver speeds in the hundreds of Mbps. However, their underwater application faces significant hurdles due to intense absorption and attenuation, which drastically reduce their effectiveness in submerged settings. Although certain frequency bands, specifically in the extremely low frequency (ELF) to very low frequency (VLF) range, exhibit relatively lower attenuation, the low data rate impedes high-speed communication. This makes it challenging for RF-based systems to maintain both a high transmission rate and the desired transmission distance underwater \cite{9449016}. 

Advancements in optical wireless communication (OWC) have broadened its scope of application, particularly in the realm of communication across water surfaces. Studies indicate that light within the blue-green spectrum, ranging from 450 nm to 550 nm, endures reasonable attenuation when transmitted through water. In contrast to traditional RF and acoustic communication methods, OWC supports high transmission rates, ensuring quick data transfer. It also features low latency, which is crucial for real-time communicationss. The combination of these unique positions OWC as a particularly promising technology for short-range communication across the water surface \cite{9114944}.

While water-to-air optical wireless communication (W2A-OWC) offers significant advantages, it also faces notable challenges that can diminish its performance in real-world scenarios. The key issues include weak signal strength, severe interference from ambient light, and the fluctuation of the communication link, which add to the complexity of deploying a robust, real-time W2A-OWC system.

In recent years, substantial research has been dedicated to the field of W2A-OWC, focusing on channel characterization and practical system implementation. Utilizing Monte Carlo simulations, the link gain and delay for W2A-OWC have been evaluated to illustrate the impact of a dynamically changing environment \cite{9209928}, \cite{8821277}. Studies \cite{9915407} have shown that minor wave actions mainly affect the energy distribution and frequency of gain variation, while larger wave movements primarily influence the coverage area. Enhancements in W2A-OWC have been achieved through the use of high-order modulation techniques and wavelength-division multiplexing, as demonstrated to improve the system's efficiency \cite{Chen:17,9592879,9115842,9184014}. In terms of system implementation, in 2019, researchers from Taipei University and National Taiwan University of Science and Technology, employed a reflective spatial light modulator with feedback to mitigate the laser beam misalignment caused by underwater turbulence. They used Pulse Amplitude Modulation (PAM) to achieve a transmission rate of up to 50 Gbps under conditions of 0.9 m underwater and 4.9 m in the atmosphere \cite{Li:19}. In 2020, a team from Tsinghua University leveraged the substantial bandwidth advantage of micro-LEDs to construct a water-to-air OWC system based on a single-layer quantum dot blue LED. A transmission rate of 2 Gbps can be achieved with a Bit Error Rate (BER) of $2.03\times10^{-3}$ across a 3 m atmospheric-underwater channel \cite{20202}.  For mid-short-range water-to-air communication, as opposed to long-distance transmission, we should prioritize enhancing the system's transmission rate, cost-effectiveness, and versatility to meet diverse communication demands while adapting to the complex underwater environment.

In this research, we have designed and implemented a real-time, multi-channel water-to-air optical wireless communication (W2A-OWC) system based on FPGA technology, which is capable of facilitating both monochromatic and polychromatic signal transmission. Such advancement allows for enhanced flexibility and application potential in various communication scenarios. The system features a configuration with multiple transmitters and receivers, which is tested in a laboratory water tank. We develop an integrated, flexible real-time system that meticulously combines several components of transmitters and receivers, including an LED array at the transmission end and avalanche photodiodes (APDs) at the reception end, into a Field-Programmable Gate Array (FPGA)-based digital signal processing system. This configuration employs the time-domain multiplexing of the LDPC decoder for multiple channels, facilitating the simultaneous transmission of multiple data streams and enhancing overall throughput. Moreover, in the developed real-time communication system, the multiplexing of  LDPC decoders can improve the utilization of resources and reduce the demand for dedicated hardware, thereby reducing system costs. To evaluate the practicality of the system, experiments are conducted in an indoor water tank, measuring the frame error rate under both calm and fluctuating water surfaces. The experimental results confirm the feasibility and effectiveness of our realization.

\section{System Model} \label{sec_system_model}
\begin{figure}[tp]
	\centerline{\includegraphics[width=1.5\columnwidth]{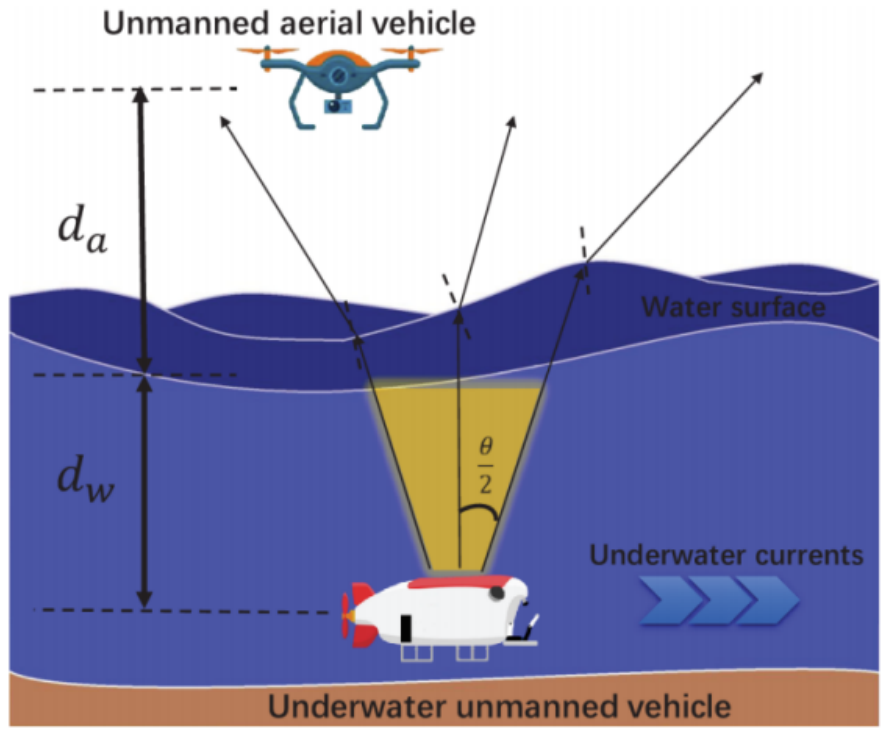}}
	\caption{ Illustration of the considered W2A-VLC system.} 
	\label{Fig1}
\end{figure}

Consider a W2A-VLC scenario where one underwater transmitter, e.g., unmanned underwater vehicle, sends information to one receiver in the air, e.g., unmanned aerial vehicle, as shown in Fig. \ref{Fig1}. The transmitter is equipped with one green/blue LED at depth  $\mathbf{d}_w$ under the water surface, and the receiver is equipped with one avalanche photodiode (APD) at height  $\mathbf{d}_a$ above the water surface. The LED with beam angle $\theta$ transmits the modulated optical signal to the APD. The light first propagates in the water and then is refracted at the wavy water surface. After propagating in the W2A link, the light is detected by the APD for further signal processing.

In this work, we establish a multi-channel  cross-water surface communication system. The transmitter consists of \textit{n} LEDs, while the receiver consists of \textit{n} APDs. Such system supports not only monochromatic light transmission but also polychromatic light transmission, covering LEDs of different wavelengths.

The APDs at the receiver are placed in one-to-one correspondence with the LEDs at the transmitting end. The output of each APD is connected to a separate amplifier and filter unit. 

\section{Hardware Implementation} \label{sec_HardwareImplementation}

We describe the components of the developed W2A-OWC real-time system, including the architectures of aerial and underwater nodes, as well as the associated software and hardware components.
\begin{figure*}[t]
	\centering  
	\includegraphics[scale=0.65]{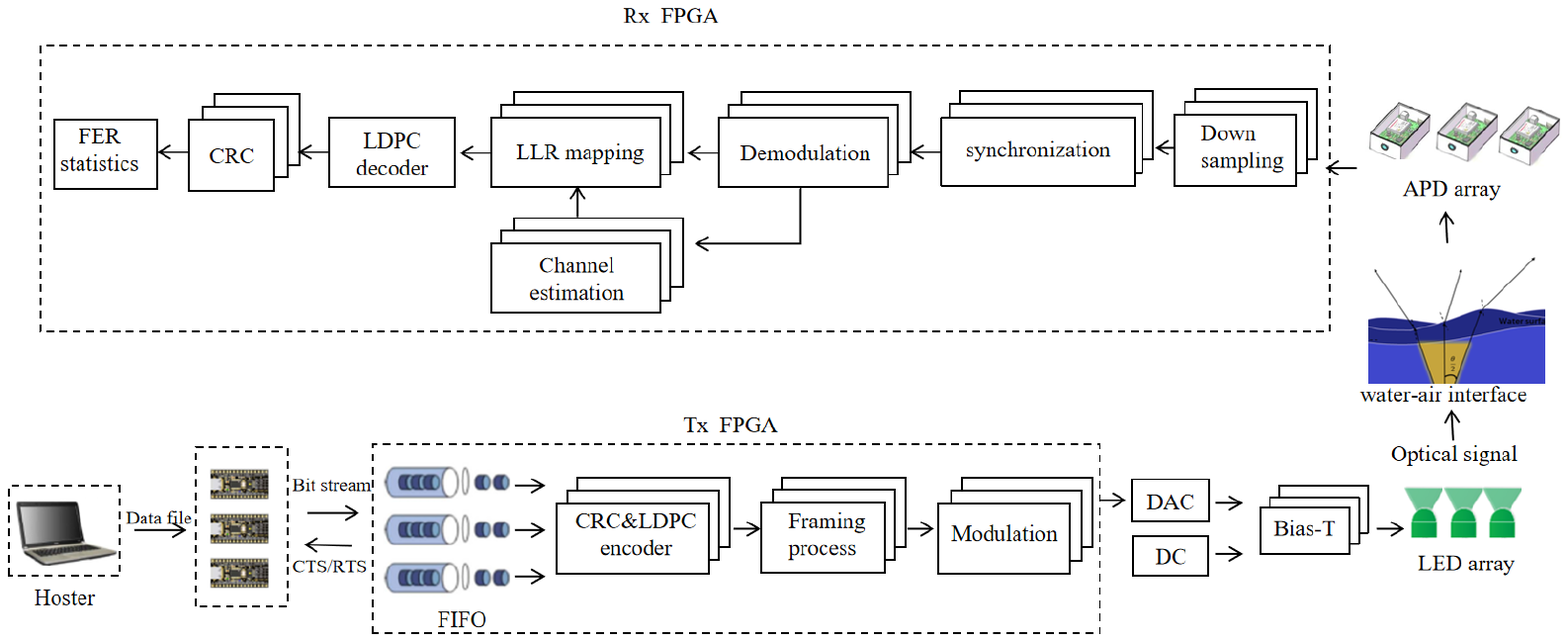}   
	\caption{The block diagram in the data processing of real-time system.}
	\label{Fig_2}
\end{figure*}

Figure~\ref{Fig_2}  shows the procedure of data processing in the real-time system, which will be introduced in two parts: the transmitter side and the receiver side.
\subsection{Transmitter Side}
At the transmitter side, the host computer's three USB interfaces utilize a USB-to-TTL module to convey three unencoded data streams to the FPGA (ALINX AX7021). Upon arrival at the FPGA, each data stream undergoes the following processes: the analog signals are sampled and transformed into digital signals, represented as "0"s and "1"s, and then stored in a First-In-First-Out (FIFO) queue. Once the data in the FIFO reaches a specified bit length, it is read out with a preceding 32-bit address information block attached, for each of the three data streams. Subsequently, the data undergoes Cyclic Redundancy Check (CRC) encoding, with a 24-bit check appended to its end, before being transmitted to the LDPC encoding IP core. The encoded data, now 2176 bits in length, is prefixed with a 256-bit synchronization sequence to form a data frame. Such data frame is then stored in the FIFO, read out according to the system's air interface rate, and modulated into an On-Off Keying (OOK) signal. Finally, the three data streams are output through the DAC (ALINX AN9767) channels as Bias-T AC signals, coupled with a DC signal from the power supply (DC = 10.8 V), to drive the three underwater node LEDs (CREE XHP70), controlling the emission of light from the LEDs. Due to the wide beam angle, the coverage area and signal reception can be enhanced, which may incur signal interference among multiple channels under too large divergence angle. Due to the limited space of the laboratory, we set the distance between each LED to 1 meter to reduce signal interference in multi-channel communication.The light signals emitted by the LED array traverse the water medium, cross the fluctuating water surface, and propagate through the atmospheric channel before ultimately reaching the receiver.
\subsection{Receiver Side}
At the receiver side, the signal light firstly reaches the photo-sensitive surface of the APDs (Hamamatsu C12702-12). The optical signals are received by three APDs and converted into three corresponding electrical signals. The receiver FPGA (ALINX AX7325B) employs an ADC (ALINX FL9627) to convert the electrical signals from the three APDs into digital signals. These digital signals pass through the blocks of downsampling, synchronization, channel estimation, and log-likelihood ratio (LLR) computation before being sent to the LDPC decoder. Afterward, the address information of the decoded output data is extracted. If the received data frame address matches the sending address, the decoded data from the three channels are separated, followed by CRC verification. Based on the CRC verification results, the count of successful decoding data frames in the register is updated. The number of received data frames can be determined by the synchronized frame count.

\textit{Remark 1:} The power consumption of the FPGA implementation of our real-time system depends on the specific FPGA model, the complexity of the implemented algorithms, and the operational conditions. In our implementation, the on-chip power consumptions of the transmitter and receiver FPGA are  0.732 W and 0.77 W, respectively. The majority of FPGA power consumption comes from dynamic power, including the clock, signal, and logic components.

\textit{Remark 2:} Regarding the resource utilization at the receiver side, the LDPC decoding core occupies 25.6$\%$ of the Slice LUT resources, with the rest accounting for 44.9$\%$. If three LDPC decoder IP cores are generated simultaneously to process these three data streams, the FPGA resources would be insufficient.

\begin{figure}[b]
	\centerline{\includegraphics[width=1\columnwidth]{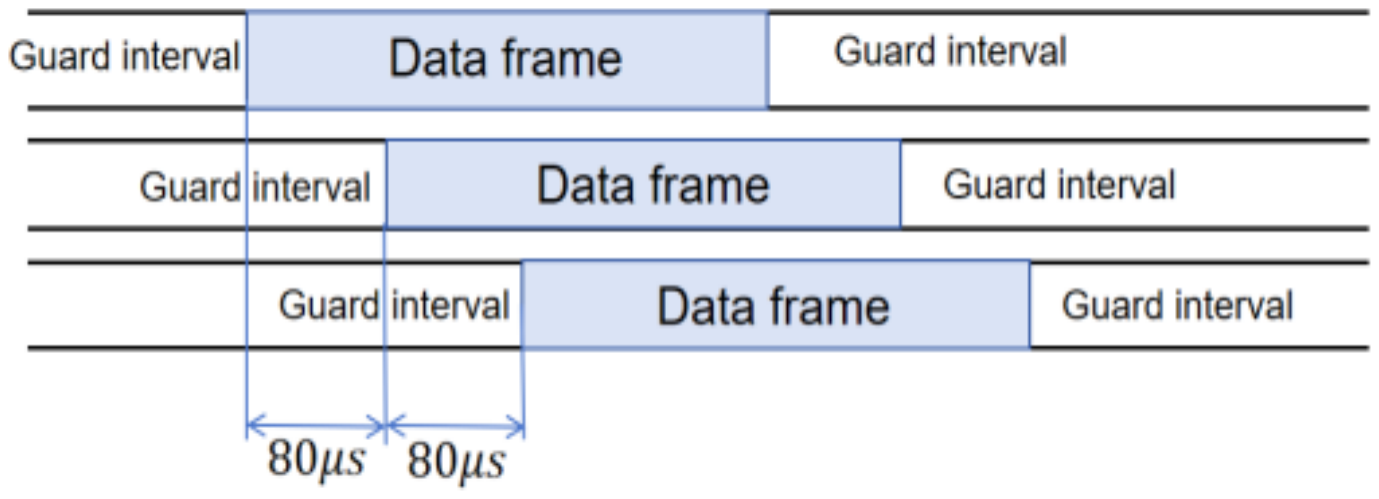}}
	\caption{The timing sequence for the transmission of the three data streams.} 
	\label{Fig_3}
\end{figure}

Next, we will explain Temporal multiplexing of the IP cores of the LDPC decoder in detail. In order for a single LDPC IP core to process three data streams, the data frames of the three data streams at the transmitting side are not transmitted simultaneously. Instead, the second stream is delayed by 80 microseconds to the first, and the third stream is delayed by another 80 microseconds to the second. When decoding with LDPC, to save resources, the three data streams are processed sequentially. After decoding, the three data streams are separated by address information, followed by CRC verification and error frame statistics. This ensures that the three received data streams are processed sequentially using a single LDPC decoder IP core, achieving efficient utilization of FPGA resources. The timing sequence for the transmission of the three data streams is shown in Fig.~\ref{Fig_3}.

\section{Experimental Results} \label{sec_experimentalresults}

\begin{figure}[tb]
	\centerline{\includegraphics[width=1.0\columnwidth]{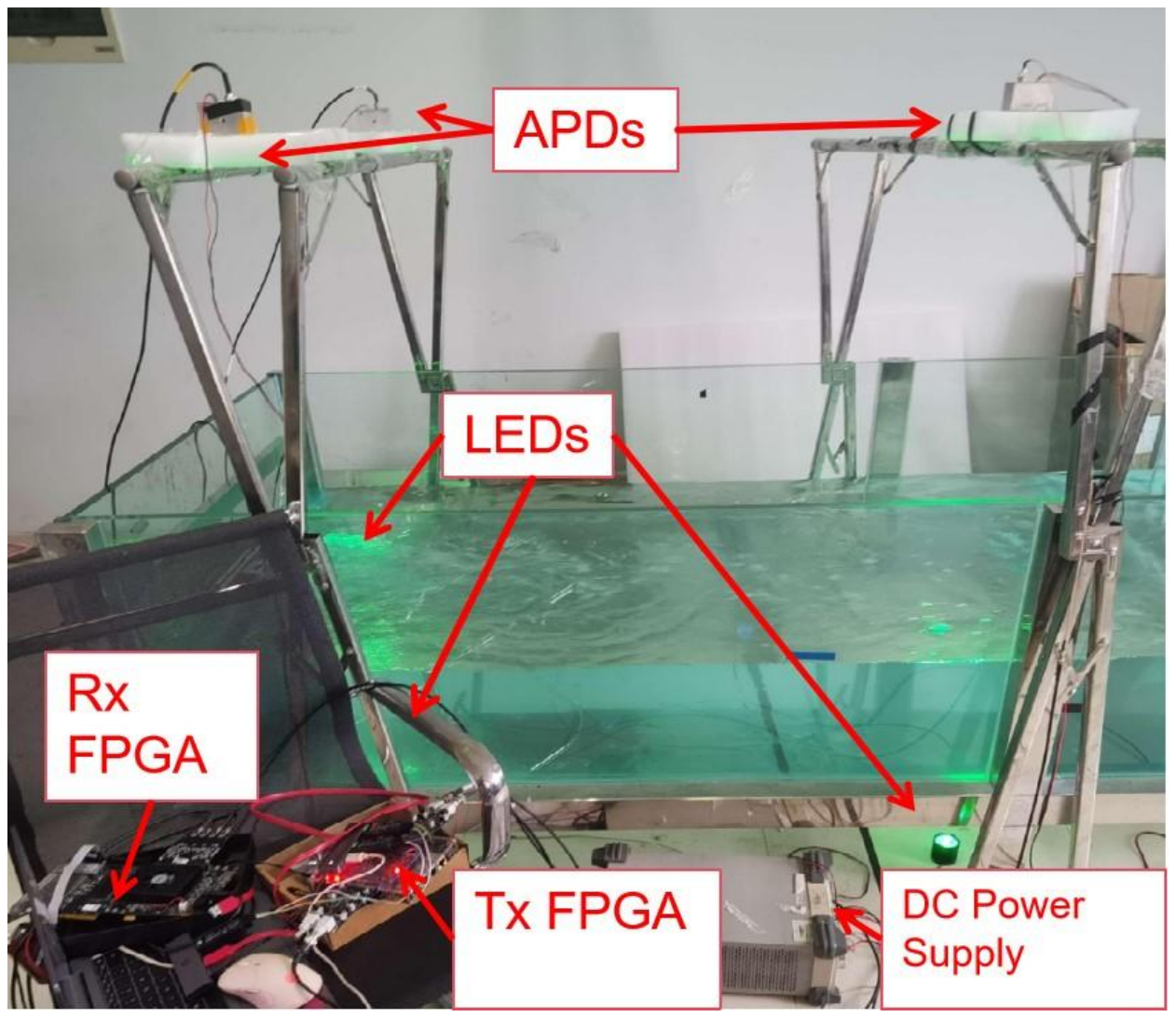}}
	\caption{ The real scene of experiment setup.} 
	\label{Fig4}
\end{figure}

\begin{table}[b]
	\caption{Experimental Parameters in Laboratory Tank}
	\centering
	\begin{tabular}{lll}
		\hline
		& Parameter                              & Value              \\ \hline
		& Coding scheme                          & 5G-NR LDPC, BG2         \\
		& LDPC rate                              & 0.588                     \\                
		& Length of data bits    & 1280               \\
		System	& Decoding algorithm                     & Min-sum algorithm  \\
		& Sampling rate                          & 100 MSPS           \\
		& Air interface rate                     & 5 Mbps             \\
		& Serial port baud rate                  & Tx: 1953125 Baud/s \\\hline		                                        
		& Depth under the water surface          & 0.47 m                \\
		\multirow{2}{*}{Transmitter} & Number of LEDs                         & 3                     \\
		& LED array configuration                & triangle vertices    \\
		& LED spacing                            & 1 m    \\\hline
		& Height above the water surface         & 0.8 m                \\
		Receiver & Number of APDs                         & 3                  \\		
		& APD array configuration                & triangle vertices    \\ \hline
		& Background radiation intensity         & 40 mW             \\
		& Attenuation coefficient                & 0.16 m$^{-1}$      \\ \hline
	\end{tabular}
\end{table}
The experimental environment in a laboratory water tank of our real-time system is shown in Fig.~\ref{Fig4}. The transmitter consists of $M=3$ LEDs for transmission, which are managed by an FPGA and a host computer. The receiver, located on a shelf, consists of $N=3$ APDs for reception, coordinated by the FPGA and host computer. The wave maker is employed to simulate disturbances in real W2A-OWC scenarios. The water tank is 3 m in long, 1 m in wide, and 0.6 m in high. The LED is placed on the ground, 0.2 m from the bottom of the tank. The water depth inside the tank is maintained at 0.27 m. The receiver is mounted on a shelf, 0.8 m above the water surface. The attenuation coefficient of the tap water in the experiment is 0.16 m$^{-1}$, slightly exceeding that of clean ocean waters. Based on the scale on the water tank wall, the wavelength of the waves ranges from 2 cm to 8 cm, and the wave height ranges from 0.5 cm to 2 cm, approximately. For a data link, the complete system parameters are shown in Table I. Rate-0.588 code is employed for transmission.

To simulate the underwater dynamic environment, the wave generator is placed in the underwater channel to make the channel dynamic. System performance is tested under calm and wavy water surface conditions.

\begin{table}[tb]
	\caption{The Experimental Results under Calm Water Surface Conditions}
	\centering
	\begin{tabular}{|c|c|c|c|}			
		\hline
		\multirow{2}{*}{Path} & \multirow{1}{*}{Number of} &  \multirow{1}{*}{Number of}   & \multirow{1}{*}{Number of frames}                \\ 	
		& \multirow{1}{*}{frames sent} &\multirow{1}{*}{frames received}&\multirow{1}{*}{with correct CRC check}\\\hline	\renewcommand{\arraystretch}{1.4}
		First  & 196633 & 196633  & 196633 \\ \hline
		Second  & 196633 & 196633 & 196633 \\ \hline
		Third  & 196633  & 196633 & 196633 \\ \hline
		
	\end{tabular}
\end{table}

\begin{table}[t]
	\caption{The Experimental Results under Wavy Water Surface Conditions}
	\centering
	\begin{tabular}{|c|c|c|c|}			
		\hline
		\multirow{2}{*}{Path} & \multirow{1}{*}{Number of} &  \multirow{1}{*}{Number of}   & \multirow{1}{*}{Number of frames}                \\ 
		& \multirow{1}{*}{frames sent} &\multirow{1}{*}{frames received}&\multirow{1}{*}{with correct CRC check}\\\hline	\renewcommand{\arraystretch}{1.4}
		First  & 233803 & 233803  & 233795 \\ \hline
		Second  & 233803 & 233803 & 233779 \\ \hline
		Third  & 233803 & 233803 & 233801 \\ \hline
		
	\end{tabular}
\end{table}

In the experiment, we employ an oscilloscope to measure the voltage of the Bias-T AC signal generated by the DAC. Table II shows the number of frames sent by the transmitter, the number of frames received, and the number of frames with correct CRC checks under calm water surface conditions at the receiver side. It can be observed that under calm water surface conditions, the average frame error rate (FER) for the three data paths is 0$\%$. Table III presents the number of frames sent by the 3 paths, the number of frames received, and the number of frames with correct CRC checks under wavy water surface conditions at the receiver end. It can be observed that under wavy water surface conditions, the average FER is $4.847\times10^{-5}$, when the voltage of Bias-T AC signals is 9.06675 Vpp in the experiment. 
\begin{figure}[tb]
	\centerline{\includegraphics[width=1.1\columnwidth]{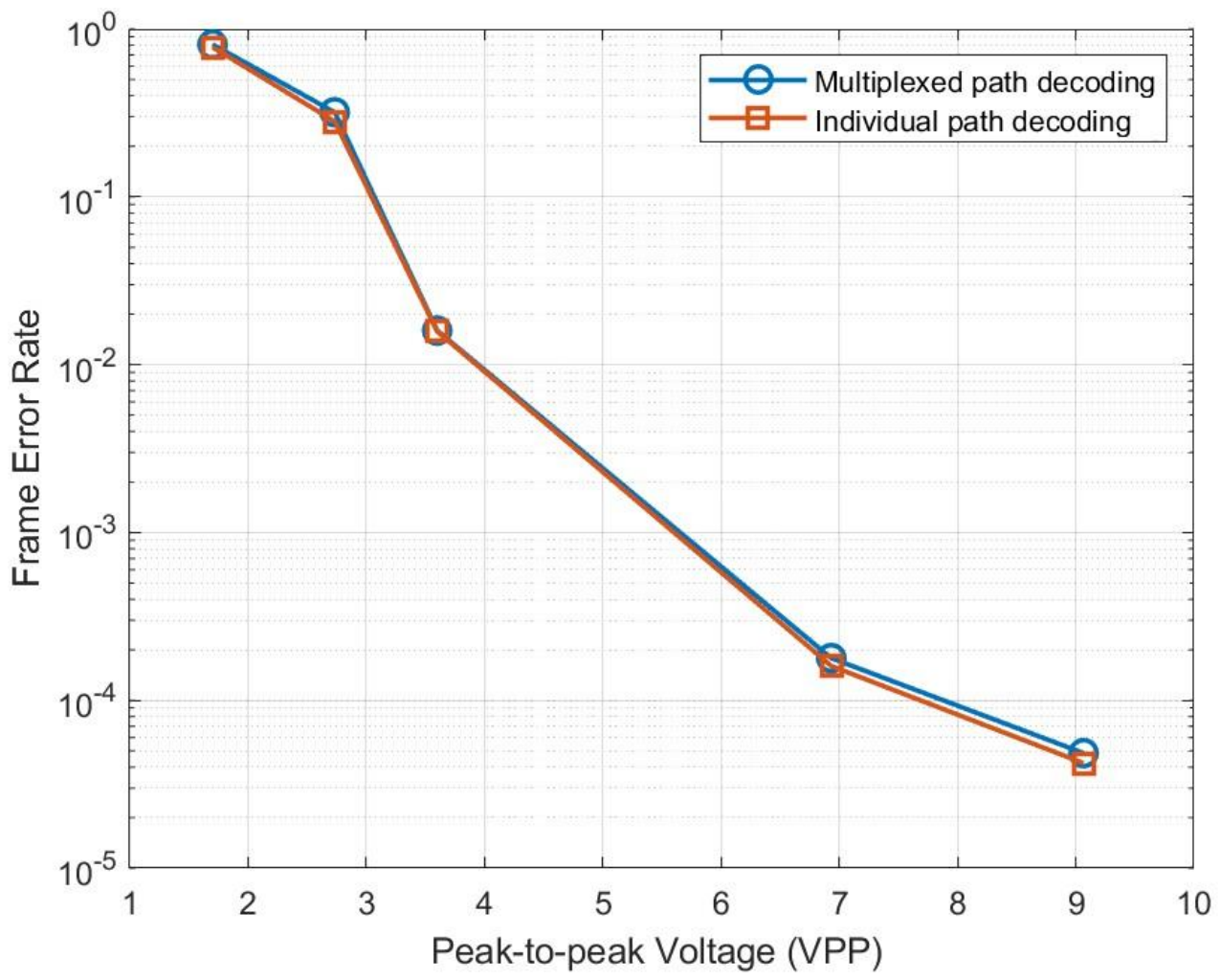}}
	\caption{The average frame error rate versus the peak-to-peak voltage for the multiplexed path decoding and the individual path decoding.} 
	\label{Fig5}
\end{figure}

To demonstrate the practicality of our system's LDPC IP core multiplexing technique, we compare the average FER when processing data individually on the three paths versus the average FER when these paths are multiplexed through the LDPC IP core according to our scheme, under varying peak-to-peak voltage of the LED and wavy water conditions. To manage data from three distinct paths, we activate one LED at a time to create a single communication link and utilize the LDPC IP core to handle a single data stream for the experiment. As shown in Fig.~\ref{Fig5}, the average FER across different peak-to-peak voltages remains relatively consistent between the two experimental approaches. Our findings suggest that our scheme is capable of effectively facilitating dynamic, real-time multi-channel visible light communication over water channels with a manageable level of complexity. Furthermore, it is observed that lower peak-to-peak voltage of the LED leads to higher FER.

\begin{figure}[tb]
	\centerline{\includegraphics[width=1.1\columnwidth]{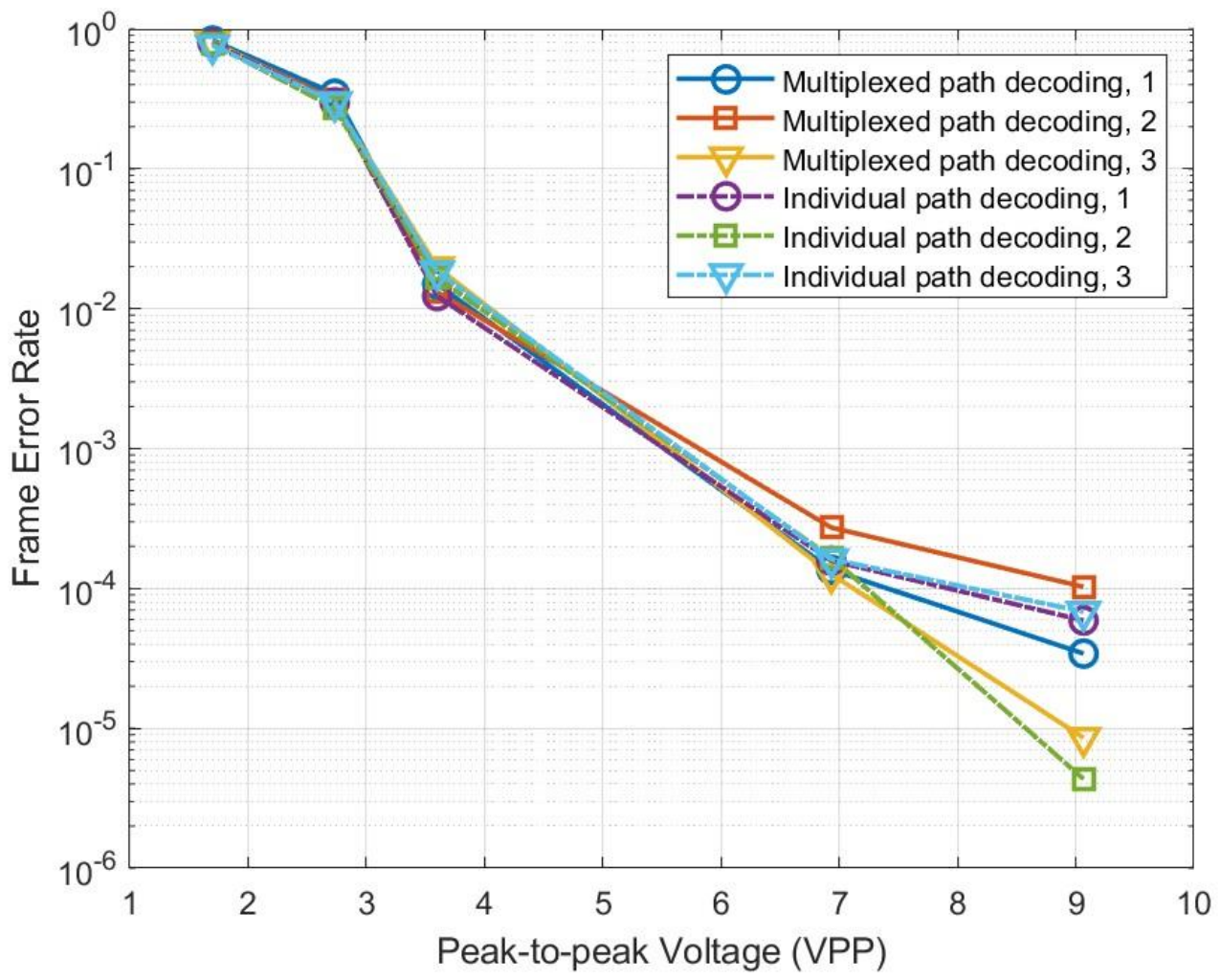}}
	\caption{The frame error rate versus the peak-to-peak voltage for the multiplexed paths and the individual paths.} 
	\label{Fig6}
\end{figure}

Figure~\ref{Fig6} presents the FERs for the three channels of our scheme, which are multiplexed through the LDPC IP core, as well as the FERs for the three channels when processing data separately. Such comparison is made under varying AC power of the LED and wavy water conditions. It can be observed that the FERs for all six paths are nearly identical under the same optical power conditions. The above results suggest that the differences in the experimental links is negligible, and confirms the effectiveness of our scheme.

\section{Conclusion} \label{sec_conclusion}
In this work, we have realized an FPGA real-time prototype of multi-channel W2A-OWC system. The real-time system utilizes 5G-NR LDPC codes to enhance data transmission reliability. Such  configuration enables parallel transmission of multiple channels, facilitating the simultaneous transmission of multiple data streams and enhancing overall throughput. Time-sharing operation of an LDPC decoder core is developed to reduce the realization cost. To evaluate the system practicality, experiments were conducted under background radiation in an indoor water tank, measuring the frame error rate under both calm and fluctuating water surfaces. The W2A-OWC system can achieve low-error transmission under wavy water surface. Experimental results confirm the feasibility and effectiveness of the developed W2A-OWC system. The system also optimizes FPGA resource usage through time-multiplexing of the LDPC decoder's IP core.
	\bibliographystyle{IEEEtran}
	\bibliography{FPGA}

\end{document}